\documentclass[11pt]{article}
\usepackage{epsfig}
\usepackage{amsmath}
\usepackage{amsfonts}
\usepackage{amstext}

\newenvironment{mysummary}[1]{%
    \leftskip=2.5em \rightskip=2.5em
    \noindent\small{\bfseries #1}}
    {\par\medskip}
\renewenvironment{abstract}{\begin{mysummary}{Abstract:}}{\end{mysummary}}
\newenvironment{keywords}{\begin{mysummary}{Key words:}}{\end{mysummary}\medskip}

\newtheorem{theorem}{\indent Theorem}

\newtheorem{definition}[theorem]{\indent Definition}

\title{Title of Article
\thanks{We have benefited from the assistance of our colleagues,
in particular ..., we also thank ... for their cooperations.}}
\author{Name1\thanks{E-mail:{\it\ ???@mail.com}},\quad Name2\quad and\quad Name3\\[1ex]
Department of ???, AMSS, CAS\\
Canberra 2600, Australia}
\date{9 January, 2011}

\begin{document}
%
\title{Quantum control theory and applications: A survey
\thanks{This paper has been published by IET Control Theory \& Applications, vol. 4, no. 12, pp.2651-2671.}
\thanks{This work was supported by
the Australian Research Council and was supported in part by the
National Natural Science Foundation of
China under Grant No. 60703083.}}

\author{Daoyi Dong \thanks{D. Dong is with the School of Engineering and Information Technology, University of New South Wales at the Australian Defence Force
Academy, Canberra, ACT 2600, Australia (email:
daoyidong@gmail.com).}, \quad Ian R. Petersen\thanks{I. R. Petersen
is with the School of Engineering and Information Technology,
University of New South Wales at the Australian Defence Force
Academy, Canberra, ACT
2600, Australia (email: i.r.petersen@gmail.com).} }

\maketitle

\begin{abstract}
This paper presents a survey on quantum control theory and
applications from a control systems perspective. Some of the basic
concepts and main developments (including open-loop control and
closed-loop control) in quantum control theory are reviewed. In the
area of open-loop quantum control, the paper surveys the notion of
controllability for quantum systems and presents several control
design strategies including optimal control, Lyapunov-based
methodologies, variable structure control and quantum incoherent
control. In the area of closed-loop quantum control, the paper
reviews closed-loop learning control and several important issues
related to quantum feedback control including quantum filtering,
feedback stabilization, LQG control and robust quantum control.
\end{abstract}

\begin{keywords}
quantum control, controllability, coherent control, incoherent
control, feedback control, robust control.
\end{keywords}

\section{Introduction}

Quantum control theory is a rapidly evolving research area, which
has developed over the last three decades \cite{Butkovskii and
Samoilenko 1979a}-\cite{Rabitz et al 2000}. Controlling quantum
phenomena has been an implicit goal of much quantum physics and
chemistry research since the establishment of quantum mechanics
\cite{Warren et al 1993}, \cite{Chu 2002}. One of the main goals in
quantum control theory is to establish a firm theoretical footing
and develop a series of systematic methods for the active
manipulation and control of quantum systems \cite{Mabuchi and
Khaneja 2005}. This goal is nontrivial since microscopic quantum
systems have many unique characteristics (e.g., entanglement and
coherence) which do not occur in classical mechanical systems and
the dynamics of quantum systems must be described by quantum theory.
Quantum control theory has already attained significant successes in
physical chemistry~\cite{Rabitz et al 2000}-\cite{Dantus 2004},
atomic and molecular physics~\cite{Chu 2002}, \cite{Bonacic 2005}
and quantum optics \cite{Wiseman and Milburn 1993}, \cite{van Handel
et al 2005-2}, and has also contributed to the understanding of
fundamental aspects of quantum mechanics \cite{Lloyd 2000}. In
recent years, the development of the general principles of quantum
control theory has been recognized as an essential requirement for
the future application of quantum technologies \cite{Dowling and
Milburn 2003}. Quantum control theory is drawing wide attention from
different communities in the areas of physics, chemistry, control
theory, applied mathematics and quantum information, and has become
a regular topic in international conferences such as the IEEE
Conference on Decision and Control. An international workshop on
``Principles and Applications for Control of Quantum Systems
(PRACQSYS)" has also been sponsored since 2004 to discuss recent
developments in this area \cite{Mabuchi and Khaneja 2005}. This
paper will present the basic concepts and main developments of
quantum control theory from a control systems perspective.

In much of quantum control theory, the controllability of quantum
systems is a fundamental issue. This issue concerns whether one can
drive a quantum system to a desired state~\cite{Huang et al 1983}.
This problem has practical importance since it has a close
connection with the universality of quantum
computation~\cite{Ramakrishna and Rabitz 1996} and the possibility
of achieving atomic or molecular scale
transformations~\cite{Ramakrishna et al 1995}, \cite{Wu et al 2006}.
Different notions of controllability, such as pure state
controllability, complete controllability, eigenstate
controllability and kinematic controllability, have been
proposed~\cite{Ramakrishna et al 1995}-\cite{Wu et al 2007}. A
common research focus is on finite dimensional (finite level)
quantum systems for which the controllability criteria may be
expressed in terms of the structure and rank of corresponding Lie
groups and Lie algebras \cite{D'Alessandro 2007}. This method allows
for an easy mathematical treatment of the problem for the case of
closed quantum systems (i.e., isolated systems considered without
external influences) and in some cases, methods from classical
(non-quantum) control theory can be directly applied. However, the
relevant criteria may be computationally difficult when the
dimension of the quantum system to be controlled is large.
Therefore, a controllability method based on graph theory has been
proposed, and the corresponding controllability criterion is easier
to verify~\cite{Turinici and Rabitz 2001}, \cite{Turinici and Rabitz
2003}. Compared with finite dimensional closed quantum systems, the
controllability of infinite dimensional quantum systems or open
quantum systems (i.e., systems considered interacting with the
environment; e.g., see \cite{Breuer and Petruccione 2002}) is more
difficult and only a few results have been obtained \cite{Wu et al
2006}, \cite{Wu et al 2007}, \cite{Altafini 2003}.

For a controllable quantum system, it is desirable to develop a good
control strategy to accomplish the required control tasks on the
quantum system \cite{Rabitz et al 2000}. The coherent control
strategy is a widely used method where one manipulates the states of
a quantum system by applying semiclassical potentials in a fashion
that preserves quantum coherence \cite{Lloyd 2000} (a wave-like
property of quantum systems allowing for constructive and
destructive interference, e.g., see \cite{Warren et al 1993}). An
early paradigm of quantum control was open-loop coherent control
\cite{Warren et al 1993}, which has successfully been used in the
quantum control of chemical reactions \cite{Rice 2001},
\cite{Shapiro and Brumer 2006}. Rigorous optimal control theory has
been successfully applied to the design of open-loop coherent
control strategies in order to find the best way of achieving given
quantum control objectives in physical chemistry \cite{Shapiro and
Brumer 2003}, \cite{Peirce et al 1988}, \cite{Dahleh et al 1990}.
Recently, time-optimal control problems for spin systems have been
solved to achieve specified control objectives in minimum time
\cite{Khaneja et al 2001}, \cite{Boscain et al 2002}, \cite{Boscain
and Mason 2006}. Optimal control techniques have also been
successfully applied to multidimensional nuclear magnetic resonance
(NMR) experiments to improve the sensitivity of these systems in the
presence of relaxation \cite{Khaneja et al 2003}, \cite{Khaneja et
al 2003b}, \cite{Khaneja et al 2004}. Another useful tool in
open-loop quantum control is the Lyapunov-based control design
approach \cite{Mirrahimi et al 2005}, \cite{Kuang and Cong 2008}. In
classical control, the Lyapunov method determines the control input
as a function of the system state. Hence, it is a feedback control
design method. In quantum control, it is difficult to acquire
information about quantum states without destroying them. Hence, the
Lyapunov-based control methodology is used to first construct an
artificial closed-loop controller and then an open-loop control law
is obtained by simulation of the artificial closed-loop system. In
coherent control, the control operations consist of unitary
transformations. However, some quantum systems may not be
controllable using only coherent controls. For such uncontrollable
quantum systems, it may be possible to enhance the capabilities of
quantum control by introducing new control strategies such as
variable structure control and incoherent control (i.e., one is
allowed to destroy coherence of the quantum systems during the
control process). For example, a variable structure control method
has been introduced in \cite{Dong VSC et al 2008}, \cite{Dong VSC et
al 2010} to enhance the capabilities of quantum control for some
specific systems which are not controllable. Incoherent control
enables the effects of quantum measurements and incoherent
environments to be combined with unitary transformations to complete
quantum control tasks and enhance the performance of quantum control
systems \cite{Vilela Mendes and Man'ko 2003}-\cite{Dong et al
2008JCP}. Incoherent control strategies have also been recognized as
important techniques to assist coherent control strategies in
quantum control systems.

Although open-loop strategies using coherent or incoherent control
approaches, optimal control techniques and Lyapunov methods, have
achieved important successes, their applications have many
limitations. In classical control, closed-loop methods have many
advantages over open-loop methods, such as robustness and
reliability. For open quantum systems, coupling with uncontrollable
environments makes the introduction of noises and uncertainties
unavoidable \cite{Gardiner and Zoller 2000}. A requirement for
robustness in the presence of uncertainties and noises has been
recognized as one of the key aspects in developing practical quantum
technologies \cite{Mabuchi and Khaneja 2005}, \cite{James et al
2008}. A natural solution to this problem is to develop closed-loop
quantum control approaches. Two paradigms for closed-loop control
have been proposed: closed-loop learning control and quantum
feedback control \cite{Rabitz et al 2000}. Closed-loop learning
control involves a closed-loop operation where each cycle of the
loop is executed with a new sample, and this approach has achieved
great success in controlling quantum phenomena in chemical reactions
\cite{Rabitz et al 2000}, \cite{Judson and Rabitz 1992}. Feedback is
an essential concept in classical control theory, where information
about the state variables obtained from direct measurements or state
estimation is fed back to the system through a controller to achieve
desired performance. Quantum feedback control has been used to
improve the system performance in different tasks, including the
control of squeezed states (a typical class of nonclassical states,
e.g., see \cite{Scully and Zubairy 1997}) and entangled states (see
Section 2.1), state reduction and quantum error correction
\cite{Wiseman and Milburn 1993}, \cite{Thomsen et al
2002}-\cite{Mancini and Wiseman 2007}. The success of feedback
control is usually dependent on the acquisition of suitable feedback
information. This problem becomes more complex and subtle in quantum
control since quantum measurements unavoidably affect the states of
measured systems. Hence, research aimed at establishing a general
framework for quantum feedback is accelerating the development of
some related areas including quantum filtering theory \cite{Belavkin
1999}, \cite{Bouten et al 2007}, \cite{Bouten et al 2008}, quantum
parameter estimation \cite{Doherty and Jacobs 1999}, \cite{Geremia
et al 2003}, \cite{Stockton et al 2004}, quantum system
identification \cite{Geremia and Rabitz 2002}, \cite{Bonnabel et al
2009}, \cite{Burgarth et al 2009 identification} and quantum robust
control \cite{James et al 2008}, \cite{D'Helon and James 2006}.

This paper will survey the development of open-loop and closed-loop
methods in quantum control theory from a control systems
perspective. We will also briefly discuss some applications of
quantum control theory to quantum information technology
\cite{Nielsen and Chuang 2000}. Many successful applications of
quantum control in chemistry and physics are beyond the scope of
this paper and, for these topics, readers can refer to several
excellent books and review articles in the areas of chemical
reactions \cite{Warren et al 1993}, \cite{Rabitz et al 2000},
\cite{Rice and Zhao 2000}, \cite{Shapiro and Brumer 2003},
\cite{Dantus 2004}, \cite{Bonacic 2005}, \cite{Rice 2001},
\cite{Shapiro and Brumer 2006}, atomic physics \cite{Chu 2002}, NMR
\cite{Mabuchi and Khaneja 2005}, \cite{Vandersypen and Chuang 2004}
and quantum optics \cite{Wiseman and Milburn 2009}. The remainder of
this paper is organized as follows. Section \ref{Section2} briefly
introduces quantum states, quantum measurements and several quantum
control models. In Section \ref{Section3}, the controllability of
quantum systems and several open-loop control strategies, including
optimal control, Lyapunov methods, incoherent control and variable
structure control, are reviewed. Section \ref{Section4} presents
some results on closed-loop control of quantum systems. In
particular, in this section, closed-loop learning control is briefly
mentioned and some specific aspects of quantum feedback control are
presented. Conclusions and perspectives are given in Section
\ref{Section5}.

\section{Prerequisites}\label{Section2}
In quantum control theory, the systems to be controlled are quantum
systems whose dynamics are governed by the laws of quantum
mechanics. Quantum mechanics provides a mathematical framework for
describing the states and evolution of quantum systems. In this
section, we briefly introduce several related concepts for
developing quantum control theory, such as quantum states, quantum
measurements and quantum control models. A comprehensive
introduction to quantum mechanics can be found in many excellent
textbooks such as \cite{Dirac 1958}, \cite{Landau and Lifshitz
1999}.

\subsection{Quantum states}
In quantum mechanics, the state of a closed quantum system can be
represented by a unit vector $|\psi\rangle$ (this notation is known
as the Dirac representation \cite{Dirac 1958} and the state
$|\psi\rangle$ is also called wavefunction) in a complex Hilbert
space $\mathcal{H}$. It is worth mentioning that the ``state" in
quantum mechanics is different from that in classical systems. For a
classical system, the state usually describes some real physical
properties such as position and momentum, which are generally
observable. However, a quantum state cannot be directly observed and
also does not directly correspond to physical quantities of the
quantum system. Since the global phase of a quantum state has no
observable physical effect, we often say that the vectors
$|\psi\rangle$ and $e^{i\alpha}|\psi\rangle$ (where $i=\sqrt{-1}$
and $\alpha \in \mathbb{R}$) describe the same physical state. A
quantum state which can be represented with a unit vector
$|\psi\rangle$ is called a pure state. For example, in quantum
information theory, information is coded using two-state (two-level)
quantum systems (called qubits) and the state $|\psi\rangle$ of a
qubit can be written as
\begin{equation}
|\psi\rangle=\cos\frac{\theta}{2}|0\rangle+e^{i\varphi}\sin\frac{\theta}{2}|1\rangle
\end{equation}
where $\theta \in [0, \pi]$, $\varphi \in [0, 2\pi]$. Then
$|0\rangle$ and $|1\rangle$ correspond to the states zero and one
for a classical bit \cite{Nielsen and Chuang 2000}, \cite{Dong et al
2008QRL}.

In practical applications, the quantum systems to be controlled are
usually not simple closed systems. They may be quantum ensembles or
open quantum systems and their states cannot be written in the form
of unit vectors. In this case, it is necessary to introduce the
density operator or density matrix $\rho: \mathcal{H}\rightarrow
\mathcal{H}$ to describe quantum states of quantum ensembles or open
quantum systems. A density operator $\rho$ is positive and has trace
equal to one. Suppose that a quantum system is in an ensemble
$\{p_{j}, |\psi_{j}\rangle\}$ of pure states; i.e., in a mixture of
a number of pure states $|\psi_{j}\rangle$ with respective
probabilities $p_{j}$. The density matrix for the system is defined
as follows \cite{Nielsen and Chuang 2000}:
\begin{equation}\label{densitymitrix}
\rho\equiv \sum_{j}p_{j}|\psi_{j}\rangle\langle \psi_{j}|
\end{equation}
where $\langle \psi_{j}|=(|\psi_{j}\rangle)^{\dag}$ and
$\sum_{j}p_{j}=1$. Here, the operation $(\cdot)^{\dag}$ refers to
the adjoint. For a pure state $|\psi\rangle$, $\rho=
|\psi\rangle\langle \psi|$ and $\text{tr}(\rho^{2})=1$. If the state
$\rho$ of a quantum system satisfies $\text{tr}(\rho^{2})<1$, we
call the quantum state a mixed state.

A composite quantum system (assumed to be made up of two subsystems
$A$ and $B$) is defined on a Hilbert space
$\mathcal{H}=\mathcal{H}_A\otimes \mathcal{H}_B$, which is the
tensor product of the Hilbert spaces $\mathcal{H}_A$ and
$\mathcal{H}_B$ on which the subsystems $A$ and $B$ are defined. For
the composite quantum system, its state $\rho_{AB}$ can be described
by the tensor product of the states of its subsystems; i.e.,
$\rho_{AB}=\rho_{A}\otimes\rho_{B}$ (e.g., see \cite{Nielsen and
Chuang 2000}).

Considering any bipartite pure state $|\psi\rangle_{AB}$, if it can
be written as a tensor product of pure states
$|\varphi\rangle_{A}\in \mathcal{H}_{A}$ and $|\chi\rangle_{B}\in
\mathcal{H}_{B}$,
\begin{equation}
|\psi\rangle_{AB}=|\varphi\rangle_{A}\otimes |\chi\rangle_{B},
\end{equation}
we call it a separable state; otherwise, we call it an entangled
state \cite{Preskill 1998}. Quantum entanglement is a uniquely
quantum mechanical phenomenon which plays a key role in many
interesting applications of quantum communication and quantum
computation, and a detailed discussion can be found in the
references \cite{Nielsen and Chuang 2000} and \cite{Preskill 1998}.

\subsection{Quantum measurements}
\subsubsection{Projective measurement}
To better control a quantum system, it is often desirable to extract
information from the controlled quantum system by means of
measurements. Measurement theory in quantum mechanics is essentially
different from that in classical mechanics since a measurement on a
quantum system unavoidably affects the measured system (a detailed
discussion of this issue can be found in \cite{Breuer and
Petruccione 2002}). In quantum mechanics, an observed physical
quantity (called an observable) is expressed as a Hermitian operator
on the Hilbert space $\mathcal{H}$ and one cannot simultaneously and
precisely measure two noncommutative observables according to the
Heisenberg uncertainty principle. A widely used measurement model is
projective measurement (or von Neumann measurement). For an
observable $M$, one can select a complete set of orthogonal
projectors $\{P_m: \sum_{m}P_{m}=I, P_{m}=P_{m}^{\dagger},
P_{\tilde{m}}P_{m}=\delta_{\tilde{m},m}P_{m}\}$ to describe a
projective measurement, where $P_{m}$ is the projector onto the
eigenspace of $M$ with eigenvalue $m$, $M=\sum_{m}mP_{m}$ and
$\delta_{\tilde{m},m}$ is the Kronecker delta \cite{Nielsen and
Chuang 2000}. For a quantum system in the state $|\psi\rangle$, the
measurement outcome will correspond to one of the eigenvalues $m$ of
the observable $M$. The probability of occurrence of outcome $m$ is
$p(m)=\langle \psi|P_{m}|\psi\rangle$. Once the outcome $m$ has
occurred, the state of the measured system changes to
$\frac{P_{m}|\psi\rangle}{\sqrt{p(m)}}$ (this process is known as
the collapse postulate). For a mixed state $\rho$, the probability
of outcome $m$ is $p(m)=\text{tr}\{P_{m}\rho\}$ and the state is
changed into $\frac{P_{m}\rho P_{m}}{p(m)}$ if we obtain the outcome
$m$.

\subsubsection{Continuous measurement}
Projective measurements are treated as instantaneous and this model
is reasonable when the strength of the measurement (i.e., the
coupling strength between the measurement apparatus and the measured
system) is large enough and the measurement time-scale is much
shorter than all other related time-scales for a given task
\cite{Jacobs 2004}. However, this framework may not be sufficient to
describe some important situations such as continuously monitoring
some aspects of a quantum system \cite{Brun 2002}, \cite{Jacobs and
Steck 2006}. In quantum feedback control, it is important to
continuously extract feedback information to adjust the system
evolution. Hence, there is an essential requirement to develop a
continuous measurement theory for quantum feedback control.
Fortunately, recent results have shown that continuous measurements
are experimentally realizable for practical quantum systems such as
solid-state qubits \cite{Korotkov 2001}. In continuous measurements,
one can continuously monitor an observable of a quantum system and
the evolution of the system described in terms of the measurement
record can be obtained from a stochastic master equation
\cite{Jacobs and Steck 2006}. Continuous measurements can be derived
from projective measurements under appropriate assumptions. For
example, to implement a continuous measurement on an atom, we let
the atom interact with an auxiliary quantum system (e.g., an
electromagnetic field), and then make projective measurements on the
auxiliary system (e.g., using a photodetector). In a short time
interval, we extract only partial information about the atom. Hence,
we also refer to continuous measurement as weak measurement or
continuous weak measurement. For a detailed introduction to
continuous measurement the reader can refer to \cite{Jacobs and
Steck 2006}. Also, the stochastic master equation which describes a
quantum system under continuous measurement will be presented in the
following subsection.

\subsection{Quantum control models}
This subsection will introduce four types of models used in quantum
control: bilinear models (BLM), Markovian master equations (MME),
stochastic master equation models (SME) and linear quantum
stochastic differential equations (LQSDE).

\subsubsection{Bilinear models}
The state $|\psi(t)\rangle$ of a closed quantum system evolves
according to the Schr\"{o}dinger equation \cite{Landau and Lifshitz
1999}
\begin{equation}\label{schrodinger}
i\hbar\frac{\partial}{\partial
t}|\psi(t)\rangle=H_{0}|\psi(t)\rangle,\qquad
|\psi(t=0)\rangle=|\psi_{0}\rangle
\end{equation}
where $H_{0}$ is the free Hamiltonian of the system (i.e., a
Hermitian operator on $\cal H$), $\hbar$ is the reduced Planck's
constant (we assume that $\hbar=1$ in this paper), and the initial
state has unit norm $\|\psi_0\|^2\equiv
\langle\psi_{0}|\psi_{0}\rangle=1$. For simplicity, we consider
finite dimensional quantum systems, which are appropriate
approximations in many practical situations. For an $N$-dimensional
quantum system, $\cal H$ is an $N$-dimensional complex Hilbert
space, and the eigenstates $\{|\phi_{i}\rangle\}_{i=1}^{N}$
(denoting $\widetilde{D}=\{|\phi_{i}\rangle\}_{i=1}^{N}$) of $H_0$
form an orthogonal basis for $\cal H$. In many situations, the
control of the system may be realized by a set of control functions
$u_{k}(t)\in \mathbb{R}$ coupled to the system via time-independent
Hermitian interaction Hamiltonians $H_{k}$ ($k=1,2,\dots$). The
total Hamiltonian $H(t)=H_{0}+\sum_{k}u_{k}(t)H_{k}$ then determines
the controlled evolution
\begin{equation}\label{BLM1}
i\frac{\partial}{\partial t}|\psi(t)\rangle
=[H_{0}+\sum_{k}u_{k}(t)H_{k}]|\psi(t)\rangle.
\end{equation}

The goal in a typical quantum control problem defined on the system
(\ref{BLM1}) is to find a final time $T>0$ and a set of admissible
controls $u_{k}(t)\in \mathbb{R}$ which drives the system from the
initial state $|\psi_{0}\rangle$ into a predefined target state
$|\psi_{f}\rangle$. The total Hamiltonian
$H(t)=H_{0}+\sum_{k}u_{k}(t)H_{k}$ defines a unitary transformation
(propagator) $U(t)$ which can accomplish the transition from the
pure state $|\psi_0\rangle$ to the pure state $|\psi(t)\rangle$;
i.e.,
\begin{equation}\label{unitary}
|\psi(t)\rangle=U(t)|\psi_0\rangle .
\end{equation}
Substituting (\ref{unitary}) into (\ref{BLM1}), we can easily obtain
\begin{equation}\label{BLM2}
i\dot{U}(t)=[H_{0}+\sum_{k}u_{k}(t)H_{k}]U(t), \ \ \ \   U(0)=I .
\end{equation}

According to the quantum state superposition principle \cite{Dirac
1958}, the evolving state $|\psi(t)\rangle$ can be expanded in terms
of the eigenstates in the set $\widetilde{D}$:
\begin{equation}\label{superposition}
|\psi(t)\rangle=\sum_{j=1}^{N}c_{j}(t)|\phi_{j}\rangle .
\end{equation}
Substituting (\ref{superposition}) into (\ref{BLM1}), we let
$C(t)=\{c_{1}(t), c_{2}(t), \dots, c_{N}(t)\}$ and obtain
\begin{equation}\label{BLM3}
i\dot{C}(t)=[H_{0}+\sum_{k}u_{k}(t)H_{k}]C(t), \ \ \ \ \
C_{0}=(c_{0j})_{j=1}^{N}, \ \ \
c_{0j}=\langle\phi_{j}|\psi_{0}\rangle .
\end{equation}

Equations (\ref{BLM1}), (\ref{BLM2}) or (\ref{BLM3}) are all
referred to as finite dimensional bilinear models (BLM) of quantum
control systems. If Equations (\ref{BLM1}), (\ref{BLM2}) and
(\ref{BLM3}) are all controllable (see Section 3.1), conversion
between them is easily carried out. For the BLM (\ref{BLM2}), we
find a set of admissible controls to generate a desired unitary
transformation $U(t)$ and then calculate the desired state
trajectory using (\ref{unitary}). For the BLM (\ref{BLM3}), we find
a set of admissible controls to achieve desired coefficient
trajectories $C(t)$ and then obtain the desired state trajectory
using (\ref{superposition}).

Bilinear models are widely used to describe closed quantum control
systems such as molecular systems in physical chemistry and spin
systems in NMR. For example, consider a spin 1/2 system in a
constant magnetic field along $z$-axis and controlled by magnetic
fields along $x$-axis and $y$-axis \cite{D'Alessandro and Dahleh
2001}. We denote the Pauli matrices
$\sigma=(\sigma_{x},\sigma_{y},\sigma_{z})$ as follows:
\begin{equation}
\sigma_{x}=\begin{pmatrix}
  0 & 1  \\
  1 & 0  \\
\end{pmatrix}, \ \ \ \
\sigma_{y}=\begin{pmatrix}
  0 & -i  \\
  i & 0  \\
\end{pmatrix}, \ \ \ \
\sigma_{z}=\begin{pmatrix}
  1 & 0  \\
  0 & -1  \\
\end{pmatrix}.
\end{equation}
The quantum control system under consideration can be described as
follows:
\begin{equation}\label{spinmodel}
i\dot{U}(t)=[I_{z}+u_{x}(t)I_{x}+u_{y}(t)I_{y}]U(t), \ \ \ \
U(0)=I
\end{equation}
where the controls $u_{x}(t), u_{y}(t)\in\mathbb{R}$, $
I_{x}=\frac{1}{2}\sigma_{x}, \  I_{y}=\frac{1}{2}\sigma_{y}$ and
$I_{z}=\frac{1}{2}\sigma_{z} . $

\subsubsection{Markovian master equations}
If we use the density matrix $\rho(t)$ to describe the state of a
closed quantum system, the evolution equation for $\rho(t)$ can be
described by the quantum Liouville equation
\begin{equation}\label{liuwell}
i \dot{\rho}(t)=[H(t), \rho(t)],
\end{equation}
where $[X, \rho]=X\rho-\rho X$ is the commutation operator. In many
practical applications, the quantum systems being considered are
open quantum systems. In fact, this is the case for most quantum
control systems since such systems unavoidably interact with their
external environments (including control inputs and measurement
devices). For an open quantum system, its evolution cannot generally
be described in terms of a unitary transformation. In many
situations, a quantum master equation for $\rho(t)$ is a suitable
way to describe the dynamics of an open quantum system. One of the
simplest cases is when a Markovian approximation can be applied
where a short environmental correlation time is supposed and memory
effects may be neglected \cite{Breuer and Petruccione 2002}. For an
$N$-dimensional open quantum system with Markovian dynamics, its
state $\rho(t)$ can be described by the following Markovian master
equation (MME) (for details, see, e.g., \cite{Breuer and Petruccione
2002}, \cite{Lindblad 1976}, \cite{Alicki and Lendi 2007}):
\begin{equation}\label{MME}
\dot{\rho}(t)=-i[H(t),\rho(t)]+\frac{1}{2}\sum_{j,k=0}^{N^{2}-1}\alpha_{jk}
\{[F_{j}\rho(t),F^{\dagger}_{k}]+[F_{j},\rho(t)F^{\dagger}_{k}]\}.
\end{equation}
Here $\{F_{j}\}_{j=0}^{N^{2}-1}$ is a basis for the space of linear
bounded operators on $\mathcal{H}$ with $F_{0}=I$, the coefficient
matrix $A=(\alpha_{jk})$ is positive semidefinite and physically
specifies the relevant relaxation rates. Markovian master equations
have been widely used as models in Markovian quantum feedback (see
Section 4.2).

\subsubsection{Stochastic master equations}
In feedback control, we generally need to continuously monitor a
quantum system to obtain feedback information. The evolution of a
quantum system under continuous measurements of an observable $X$
can be described by the following master equation \cite{Jacobs and
Steck 2006}:
\begin{equation}\label{SME1}
d\rho=-i[H, \rho]dt-\kappa[X, [X, \rho]]dt+\sqrt{2\kappa}(X\rho+\rho
X-2\langle X\rangle\rho)dW
\end{equation}
where $\kappa$ is a parameter related to the measurement strength,
$\langle X\rangle=\text{tr}(X\rho)$, $dW$ is a Wiener increment with
zero mean and variance equal to $dt$ and satisfies the following
relationship to the measurement output $y$
\begin{equation}\label{Wiener}
dW=dy-2\sqrt{\kappa}\text{tr}(X\rho)dt .
\end{equation}
Equation (\ref{SME1}) is usually called a stochastic master equation
(SME). Such a stochastic master equation can be obtained as a
filtering equation from quantum filtering theory (for details, see
\cite{van Handel et al 2005-2}, \cite{van Handel et al 2005}). It is
worth noting that the state $\rho$ in the SME (\ref{SME1}) is a
conditional state since it uses information from the continuous
measurements. It should be pointed out that (\ref{SME1}) is only a
typical form of SME and there exist many different types of SMEs
which depend on different measurement processes \cite{Qi 2009}.

Stochastic master equations have been derived for some quantum
optical systems under continuous measurements. For example, we
consider an atomic ensemble interacting with an electromagnetic
field considered in \cite{van Handel et al 2005-2}. Consider the
atomic Hamiltonian $H(t)= \Delta F_{z}+ u(t)F_{y}$, where $\Delta$
is the atomic detuning and $u(t)$ is the strength of a magnetic
field in the $y$-direction, and $F_{z}$ and $F_{y}$ are the
collective dipole moments of the ensemble (for details, see
\cite{van Handel et al 2005-2}). If we neglect spontaneous emission,
the corresponding SME is \cite{van Handel et al 2005-2}
\begin{equation}\label{SME2}
d\rho=-i [u(t)F_{y}+s F_{z}, \rho]dt-\kappa[F_{z}, [F_{z},
\rho]]dt+\sqrt{2\kappa}(F_{z}\rho+\rho F_{z}-2\langle
F_{z}\rangle\rho)dW
\end{equation}
where $s$ is related to experimental parameters such as $\Delta$,
and $dW$ satisfies
\begin{equation}\label{Wiener2}
dW=dy-2\sqrt{\kappa}\text{tr}(F_{z}\rho)dt .
\end{equation}

\subsubsection{Linear quantum stochastic differential equations}
In the BLM (\ref{BLM1}), MME (\ref{MME}) and SME (\ref{SME1}), we
use the Schr\"{o}dinger picture of quantum mechanics where equations
describing the time dependence of quantum states are given. In some
cases, it is more convenient to adopt the Heisenberg picture where
the time-dependent operators on $\mathcal{H}$ are used in describing
the quantum dynamics. An interesting case is a class of
noncommutative linear stochastic systems which includes many
examples of interest in quantum technology, especially in linear
quantum optics \cite{James et al 2008}. This class of systems can be
described by the following linear quantum stochastic differential
equation (LQSDE) \cite{James et al 2008}:
\begin{equation}\label{LQSE}
\begin{array}{l}
dx(t)=Ax(t)dt+Bdw(t); \ x(0)=x_{0} \\ dy(t)=Cx(t)dt+Ddw(t)
\end{array}
\end{equation}
where $A$, $B$, $C$ and $D$ are, respectively, $\mathbb{R}^{n\times
n}$, $\mathbb{R}^{n\times n_{w}}$, $\mathbb{R}^{n_{y}\times n}$ and
$\mathbb{R}^{n_{y}\times n_{w}}$ matrices ($n, n_{w}, n_{y}$ are
positive integers), and $x(t)=[x_{1}(t) \dots x_{n}(t)]^{T}$ is a
vector of self-adjoint possibly noncommutative system variables.

The initial system variables $x(0)=x_{0}$ consist of operators (on
an appropriate Hilbert space) satisfying the commutation relations
\begin{equation}
[x_{j}(0), x_{k}(0)]=2i \Theta_{jk}, j,k=1, \dots, n,
\end{equation}
where $\Theta_{jk}$ are the components of a real antisymmetric
matrix $\Theta$. For simplicity, we may take $\Theta$ to have one of
the following forms: (i) Canonical if $\Theta=\text{diag}(J,\dots,
J)$, or (ii) degenerate canonical if $\Theta=\text{diag}(0_{n'\times
n'}, J, \dots, J)$, where $0<n'\leq n$, $n'$ is the number of
classical variables, and $J=[\begin{smallmatrix}0 && 1
\\ -1 && 0\end{smallmatrix}]$. The vector quantity $w$
describes the input signals and is assumed to admit the
decomposition
\begin{equation}\label{noiseinput}
dw(t)=\beta_{w}(t)dt+d\tilde{w}(t)
\end{equation}
where $\tilde{w}(t)$ is the noise part of $w(t)$ and $\beta_{w}(t)$
is a self-adjoint, adapted process (see \cite{Parthasarathy 1992},
\cite{Belavkin 1992}). The noise $\tilde{w}(t)$ is a vector of
self-adjoint quantum noises with It\^{o} table
\begin{equation}
d\tilde{w}(t)d\tilde{w}^{T}(t)=F_{\tilde{w}}dt
\end{equation}
where $F_{\tilde{w}}$ is a positive semidefinite Hermitian matrix
(see \cite{Parthasarathy 1992}, \cite{Belavkin 1992} for details).
Equation (\ref{LQSE}) can describe quantum systems such as linear
quantum optical systems. However, it does not necessarily represent
the dynamics of a meaningful physical system. We may need to add
some additional constraints to ensure that a system described by
equations of the form (\ref{LQSE}) is physically realizable (for
details, see Section \ref{Section4} or \cite{James et al 2008}).

\section{Controllability and open-loop control of quantum systems}\label{Section3}
\subsection{Controllability}
The controllability of quantum systems is a fundamental theoretical
notion in quantum control as well as having practical importance
because of its close connection with the universality of quantum
computation~\cite{Ramakrishna and Rabitz 1996} and the possibility
of attaining atomic or molecular scale
transformations~\cite{Ramakrishna et al 1995}, \cite{Wu et al 2006}.
Different notions of controllability for quantum systems have been
proposed and some notions have been studied in
depth~\cite{D'Alessandro 2007}. In this section, we will briefly
present some aspects of controllability for quantum systems. A
common research focus is on BLM of quantum systems for which the
controllability criteria may be expressed in terms of the structure
and rank of the corresponding Lie groups and Lie
algebras~\cite{D'Alessandro 2007}. We denote the $N$-dimensional
complex unit sphere by $S^{N}_{\mathcal{C}}$ and the Lie algebra
generated by the operators $\{-iH_{0}, -iH_{1}, \dots, -iH_{K}\}$ as
$\mathcal{L}_{0}$. Also, we let $U(N)$ ($SU(N)$) represent the
$N$-dimensional unitary group (special unitary group) and $u(N)$
($su(N)$) be the corresponding Lie algebra. Furthermore, we let
$\mathcal{R}(|\psi\rangle)$ denote the reachable set of all states
that are reachable from $|\psi\rangle$. Now, we give several
definitions of controllability and their corresponding testing
criteria.

\begin{definition}[Pure State Controllability]\cite{Albertini and D'Alessandro
2003} The quantum system (\ref{BLM1}) is pure state controllable if
for every pair of initial and final states, $|\psi_{0}\rangle$ and
$|\psi_{f}\rangle$ in $S^{N-1}_{\mathcal{C}}$ there exist control
functions $\{u_{k}(t)\}$ and a time $t>0$ such that the
corresponding solution of (\ref{BLM1}) at time $t$, with initial
condition $|\psi_{0}\rangle$, is $|\psi(t)\rangle=|\psi_{f}\rangle$.
\end{definition}
\begin{definition}[Operator Controllability] \cite{Albertini and D'Alessandro
2003} The quantum system (\ref{BLM2}) is operator controllable if
there exist control functions $\{u_{k}(t)\}$ which drive the unitary
operator $U$ in (\ref{BLM2}) from $I$ to $U_{f}$, for any $U_{f}\in
U(N)$ (or $SU(N)$).
\end{definition}
\begin{definition}[Eigenstate Controllability]\cite{Zhang et al 2005}
Suppose $|\phi_{1}\rangle, \dots, |\phi_{N}\rangle$ are the $N$
eigenstates of the free Hamiltonian $H_{0}$. The quantum system
(\ref{BLM1}) is eigenstate controllable if
$\bigcup_{i=1}^{N}\mathcal{R}(|\phi_{i}\rangle)=S_{\mathcal{C}}^{N-1}$.
\end{definition}

\begin{theorem}\cite{D'Alessandro 2007}\label{PSCtest}
The quantum system (\ref{BLM1}) is pure state controllable if and
only if the corresponding dynamical Lie algebra $\mathcal{L}_{0}$
satisfies one of the following conditions: (1)
$\mathcal{L}_{0}=su(N)$; (2) $\mathcal{L}_{0}$ is conjugate to
$sp(\frac{N}{2})$, where $sp(\frac{N}{2})$ is the
$\frac{N}{2}$-dimensional symplectic group; (3)
$\mathcal{L}_{0}=u(N)$; (4)
$\mathcal{L}_{0}=\text{span}\{i\textbf{1}_{N\times N}\}\oplus
\tilde{\mathcal{L}}$, where $\tilde{\mathcal{L}}$ is a Lie algebra
conjugate to $sp(\frac{N}{2})$.
\end{theorem}

\begin{theorem}\cite{Albertini and D'Alessandro 2003}\label{OCtest}
The quantum system (\ref{BLM2}) is operator controllable if and only
if $\mathcal{L}_{0}=u(N)$ (or $\mathcal{L}_{0}=su(N)$).
\end{theorem}

From the above definitions and theorems, we can see that the notion
of operator controllability is strongest of these quantum
controllability notions and eigenstate controllability is weakest.
If a quantum system is operator controllable, it is easy to show
that it is also pure state controllable using (\ref{unitary}) or by
a comparison of Theorems \ref{PSCtest} and \ref{OCtest}. If a
quantum system is pure state controllable, it follows from the
definitions that it must also be eigenstate controllable. Note that
pure state controllability is also called wavefunction
controllability in some references (e.g., \cite{Turinici and Rabitz
2001}, \cite{Turinici and Rabitz 2003}, \cite{Dong VSC et al 2008}).
Operator controllability, in the unitary case, is also called
complete controllability \cite{Schirmer et al 2001}, \cite{Schirmer
et al 2002}. Some algorithms have been developed for testing the
controllability of specific quantum systems \cite{Ramakrishna et al
1995}, \cite{Schirmer et al 2001}, \cite{D'Alessandro 2007},
\cite{Altafini 2002}.

The Lie algebra method allows for a straightforward mathematical
treatment of closed quantum systems. However, the relevant criteria
may be computationally difficult when the dimension of the
controlled system is large. Hence, another method based on graph
theory has been developed for pure state controllability for which
the controllability criterion becomes easy to verify. For
simplicity, we consider the quantum system (\ref{BLM3}) with only a
single control input $u(t)$; i.e.,
\begin{equation}\label{WCmodel}
i\dot{C}(t)=AC(t)+u(t)BC(t) \ \ \ \ \ \ C_{0}=(c_{0j})_{j=1}^{N}
\end{equation}
where the matrices $A$ and $B$ correspond to the free Hamiltonian
and interaction Hamiltonian, respectively. Now, we associate the
system with a non-oriented connectivity graph $G=(V,E)$, where the
set of vertices $V$ corresponds to the set of the eigenstates
$|\phi_{i}\rangle$ and the set of edges $E$ corresponds to the set
of all pairs of eigenstates directly coupled by the matrix
$B=(B_{ij})$,
\begin{eqnarray}\label{graph}
G(V,E):&&V=\{|\phi_{1}\rangle,\dots,|\phi_{N}\rangle\},\nonumber\\
&&E=\{(|\phi_{i}\rangle,|\phi_{j}\rangle);i<j, B_{ij}\neq 0\} \ .
\end{eqnarray}
Let $G_{k}=(V^{(k)},E^{(k)}), k=1,\dots,K$ be connected components
of this graph. We denote by $\lambda_{i}$ $(i=1,\dots, N)$ the
eigenvalues of the matrix $A$ and let
$\nu_{ij}=\lambda_{i}-\lambda_{j}$ ($i,j=1,\dots,N$). The following
theorem provides a sufficient condition for pure state
controllability in terms of the connectivity graph~\cite{Turinici
and Rabitz 2001}, \cite{Turinici and Rabitz 2003}:

\begin{theorem}\label{WC}
The system (\ref{WCmodel}) is pure state controllable if the
following
conditions hold:\\
(I) The graph $G$ is connected; i.e. $K=1$.\\
(II) The graph $G$ does not have ``degenerate transitions". That is
for all $(i,j)\neq (a,b)$, $i\neq j$, $a\neq b$ such that
$B_{ij}\neq 0$, $B_{ab}\neq
0$, then $\nu_{ij}\neq \nu_{ab}$.\\
(III) For each $i,j,a,b=1,\dots,N$ such that $\nu_{ij}\neq 0$ the
number $(\nu_{ab}/\nu_{ij})$ is rational.
\end{theorem}
The proof of Theorem \ref{WC} can be found in~\cite{Turinici and
Rabitz 2003}. The three conditions in Theorem \ref{WC} provide a
sufficient but not a necessary condition for pure state
controllability. In some circumstances, the assumptions (II) and
(III) can be slightly relaxed~\cite{Turinici and Rabitz 2001}.
Moreover, it is clear that if we add one more eigenstate into the
set $\widetilde{D}$, the pure state controllability criterion is
easy to check for the resultant new system~\cite{Turinici and Rabitz
2001}.

All of the above results consider finite dimensional quantum
systems, which are a reasonable approximation to the quantum systems
arising in many applications. Many practical quantum systems,
especially those with continuous spectra, are essentially infinite
dimensional; i.e., $|\psi(t)\rangle$ in (\ref{BLM1}) is a unit
vector in an infinite dimensional Hilbert space $\mathcal{H}$. The
controllability of such infinite dimensional quantum systems has
been studied in \cite{Huang et al 1983}. For an infinite dimensional
system (\ref{BLM1}), the existence of a dense analytic domain (in
the sense of Nelson), together with some conditions on the
corresponding Lie algebra, can guarantee its analytic
controllability. The controllability of infinite dimensional quantum
control systems has also been extensively studied in \cite{Wu et al
2006}, \cite{Lan et al 2005}.

Another interesting focus is on the controllability of open quantum
systems. It has been proven that a finite dimensional open quantum
system with Markovian dynamics (i.e., MME (\ref{MME})) is not
controllable when using only coherent control \cite{Altafini 2003}.
However, finite dimensional open quantum systems with Kraus-map
dynamics are complete kinematic state controllable \cite{Wu et al
2007} and a specific Kraus map can be constructed for transformation
from an arbitrary initial state to a predefined target state if some
incoherent resources are available as control tools \cite{Kraus
1983}, \cite{Lloyd and Viola 2001}. A control approach involving
generating Kraus type dynamics can be potentially robust to
variations of the initial system state \cite{Wu et al 2007}.

\subsection{Optimal control}
Most results on the controllability of quantum systems do not
provide constructive methods to design a control law for driving a
quantum system from an initial state to a predetermined target
state. Hence, it is desirable to develop useful methods to design
such a control law. A straightforward strategy for constructing a
suitable control law is based on a Lie group decomposition, where
the desired system evolution operator $U_{f}$ is decomposed as the
product of basic unitary operators which are easy to generate using
simple controls \cite{D'Alessandro 2007}, \cite{Ramakrishna et al
2000}, \cite{Khaneja and Glaser 2001}. This method is constructive
and exact. However, it may be difficult to complete the
decomposition for most practical systems. For many practical quantum
control problems, optimal control theory is a powerful tool in
achieving quantum control objectives \cite{Peirce et al 1988},
\cite{Werschnik and Gross 2007}.

In the optimal control approach, the quantum control problem can be
formulated as a problem of seeking a set of admissible controls
satisfying the system dynamic equations and simultaneously
minimizing a cost functional. The cost functional may be different
according to the practical requirements of the quantum control
problem, such as minimizing the control time \cite{Khaneja et al
2001}, \cite{Sugny et al 2007}, the control energy
\cite{D'Alessandro and Dahleh 2001}, \cite{Grivopoulos and Bamieh
2008}, the error between the final state and target state, or a
combination of these requirements. Many useful tools in traditional
optimal control, such as the variational method, the Pontryagin
minimum principle and convergent iterative algorithms \cite{Zhu and
Rabitz 1998}, can be adapted to quantum systems and applied to
search for optimal controls. Optimal control techniques have been
widely applied to control quantum phenomena in physical chemistry
(for details, see, e.g., \cite{Rabitz et al 2000}, \cite{Rice and
Zhao 2000}, \cite{Shapiro and Brumer 2003}, \cite{Rabitz et al
2004}, \cite{Chakrabarti and Rabitz 2007}) and NMR experiments
\cite{Khaneja et al 2003b}, \cite{Khaneja et al 2004},
\cite{Vandersypen and Chuang 2004}.

Here we present a simple example to demonstrate optimal quantum
control design. We consider the model (\ref{spinmodel}) and assume
$u_{y}(t)=0$. We can produce an $x$ rotation of the spin by using
radio-frequency (RF) pulses to control the system. The quantum
control system becomes
\begin{equation}\label{spinxmodel}
i\dot{U}(t)=[I_{z}+u_{x}(t)I_{x}]U(t), \ \ \ \   U(0)=I
\end{equation}
From Theorem \ref{OCtest}, it is easy to check that the system
(\ref{spinxmodel}) is operator controllable (completely
controllable). It is possible to design a control law which is
optimal in some sense to accomplish a specified quantum control
task. In some applications, it is desirable to accomplish the
control task as soon as possible in order to minimize the relaxation
effects \cite{Khaneja et al 2001}. For example, it is desirable to
perform a quantum computation task with a minimum time using a
collection of control resources \cite{Yuan and Khaneja 2005}. Hence,
the time-optimal control of quantum systems is an interesting
practical problem \cite{Khaneja et al 2001}, \cite {Sugny et al
2007}. If the control is not bounded, we have the following theorem
\cite{Khaneja et al 2001}.
\begin{theorem}
Consider the model (\ref{spinxmodel}). Given any $U_{f} \in SU(2)$,
there exists a unique $\beta\in [0, 2\pi]$ such that
$U_{f}=\exp(-i\alpha I_{x})$ $\exp(-i\beta I_{z})$ $\exp(-i\gamma
I_{x})$, where $\alpha$, $\gamma \in \mathbb{R}$, and the minimum
time required in producing $U_{f}$ is $\beta$.
\end{theorem}

If we ignore the global phase, we can denote the initial state
$|\psi_{0}\rangle$ and the target state $|\psi_{f}\rangle$ as
\begin{equation}\label{initial}
|\psi_{0}\rangle=\cos{\frac{\theta_{0}}{2}}|0\rangle+e^{i\varphi_{0}}\sin{\frac{\theta_{0}}{2}}|1\rangle,
\end{equation}
\begin{equation}\label{final}
|\psi_{f}\rangle=\cos{\frac{\theta_{f}}{2}}|0\rangle+e^{i\varphi_{f}}\sin{\frac{\theta_{f}}{2}}|1\rangle
\end{equation}
where $\theta_{0}, \theta_{f} \in [0, \pi]$, $\varphi_{0},
\varphi_{f} \in [0, 2\pi]$ and
$|\psi_{f}\rangle=U_{f}|\psi_{0}\rangle$. We can give an analytical
expression for the minimal time required to accomplish this control
task using geometric manipulations on the Bloch sphere \cite{Dong et
al 2008-2}. This minimum time $\beta$ can be calculated according to
the following equation:
\begin{equation}\label{timeoptimal}
\beta=|\arccos(\sin\theta_{f}\cos\varphi_{f})-\arccos(\sin\theta_{0}\cos\varphi_{0})|
\ .
\end{equation}
A formal proof of this result can be found in \cite{Dong et al
2008-3}.

Under the assumption of bounded control (i.e., $|u(t)|\leq V\in
\mathbb{R}^{+}$), an interesting case is bang-bang control where the
control $u(t)$ switches between two values $\pm V$. The solutions to
many optimal problems take the form of bang-bang control
\cite{Boscain and Mason 2006}. The bang-bang control strategy has
been used to suppress decoherence \cite{Viola and Lloyd 1998} (the
evolution of pure states to mixed states; e.g., see \cite{Preskill
1998}) and to manipulate a class of coherent states \cite{Wu Jianwu
et al 2007}. To reduce the error introduced by the switches, the
number of switches can be used as a performance index for optimal
control design. Several algorithms have been developed to determine
the minimum number of switches required. In one approach, the number
of switches is determined by finding the minimum number of factors
required in the factorization of $U_{f}$ \cite{D'Alessandro 2002}.
Another approach involves geometric analysis on the corresponding
complex unit sphere (for details, see \cite{Dong et al 2008-3}).

From the above example, we can see that, for the quantum system
(\ref{spinxmodel}), we may wish to optimize different cost
functionals depending on the application. We may also use different
methods to solve the same optimal quantum control problem. Optimal
control is often combined with other strategies, such as closed-loop
learning control and quantum feedback control, when manipulating
quantum entanglement \cite{Mancini and Wiseman 2007}, identifying
Hamiltonian parameters \cite{Geremia and Rabitz 2002}, controlling
chemical reactions \cite{Rabitz et al 2000}, \cite{Shapiro and
Brumer 2003}, and tracking quantum states \cite{Chen et al 2008}.

\subsection{Lyapunov-based control design}
Lyapunov-based control methods are powerful tools for feedback
controller design in classical control theory. In quantum control,
the acquisition of feedback information through measurements usually
destroys the state being measured, which makes it difficult to
directly apply Lyapunov approaches to quantum feedback controller
design. However, we may first complete the feedback control design
by simulation on a computer, which will give a sequence of controls.
Then we can apply the control sequence to the quantum system to be
controlled in an open-loop form \cite{Mirrahimi et al 2005},
\cite{Kuang and Cong 2008}, \cite{Ferrante et al
2002}-\cite{Altafini 2007}. This is a ``feedback design and
open-loop control" strategy. This strategy is especially useful for
some difficult quantum control tasks \cite{Altafini 2007-1}. The
most important aspects in Lyapunov-based control design include the
construction of the Lyapunov function, the determination of the
control law and the analysis of asymptotic convergence.

To design an open-loop control law for quantum systems, several
types of Lyapunov functions have been considered. For example, if
the target state is $|\psi_{f}\rangle$, we may select one of the
following Lyapunov functions: $V_{1}(t)=\frac{1}{2}(1-|\langle
\psi_{f}|\psi(t)\rangle|^{2})$ \cite{Vettori 2002},
 $V_{2}(t)=\langle
\psi(t)-\psi_{f}|\psi(t)-\psi_{f}\rangle$ \cite{Mirrahimi et al
2005}, or $V_{3}(t)=\langle \psi(t)|P|\psi(t)\rangle$, where $P$ is
a positive semidefinite Hermitian operator \cite{Grivopoulos and
Bamieh 2003}. It is clear that $V_{j}(t)\geq 0$ ($j=1,2,3$). We can
now select the control function to guarantee that the first-order
time derivative of the Lyapunov function is negative semidefinite.
That is, we may determine the control law using the condition
$\dot{V}(t)\leq 0$ on the Lyapunov function $V(t)$.

For the case that the target state is an eigenstate of the free
Hamiltonian $H_{0}$, several algorithms have been developed to find
corresponding control laws and their convergence has been analyzed
\cite{Kuang and Cong 2008}, \cite{Vettori 2002}, \cite{Grivopoulos
and Bamieh 2003}. For the case that target states are not
eigenstates of $H_{0}$, the control problem can be formulated in
terms of reference trajectory tracking \cite{Mirrahimi et al 2005},
\cite{Mirrahimi et al 2005-2}, \cite{Wang and Schirmer 2010}.
Lyapunov-based techniques have also been formulated using density
operators for the control of a spin ensemble \cite{Altafini 2007-1},
\cite{Altafini 2007}. LaSalle's invariance principle is also useful
to characterize the asymptotic behavior of the system dynamics
\cite{D'Alessandro 2007}.

\subsection{Variable structure control}
For some quantum systems which are not controllable, we can
introduce additional resources or special strategies to change the
controllability of the quantum system or enhance our capability of
controlling the quantum system. In this subsection, we introduce a
variable structure control strategy to enhance our capability of
controlling a class of quantum systems. In variable structure
control, one can change the controller structure according to
specified switching logic in order to obtain improved properties
\cite{Dong VSC et al 2008}.

Consider an example corresponding to two special cases of
(\ref{BLM3}):
\begin{equation}\label{model1}
i\dot{C}(t)=AC(t)+u_{1}(t)B_{1}C(t)=(A+B^{u}_{1})C(t) ;
\end{equation}
\begin{equation}\label{model2}
i\dot{C}(t)=AC(t)+u_{2}(t)B_{2}C(t)=(A+B^{u}_{2})C(t)
\end{equation}
where $A=\text{diag}\{1.0, 1.2, 1.3, 2.0, 2.15\}$, and
$$
B_{1}^{u}=
\begin{pmatrix}
  0   & 0  & 0  & u_{1}(t)  & u_{1}(t)   \\
  0   & 0  & 0  & 0  & 0   \\
  0   & 0  & 0  & 0  & 0   \\
  u_{1}(t)   & 0  & 0  & 0  & u_{1}(t)   \\
  u_{1}(t)   & 0  & 0  & u_{1}(t)  & 0  \\
\end{pmatrix};\ \ \
B^{u}_{2}=
\begin{pmatrix}
  0   & 0  & u_{2}(t)  & 0  & 0   \\
  0   & 0  & 0  & u_{2}(t)  & u_{2}(t)   \\
  u_{2}(t)   & 0  & 0  & 0  & 0   \\
  0   & u_{2}(t)  & 0  & 0  & u_{2}(t)   \\
  0   & u_{2}(t)  & 0  & u_{2}(t)  & 0  \\
\end{pmatrix}
$$

In \cite{Dong VSC et al 2008}, it has been shown that the two models
(\ref{model1}) and (\ref{model2}) are not individually pure state
controllable but are controllable under a variable structure control
strategy. Furthermore, consider a system ($S_m$):
\begin{equation}
i\dot{C}(t)=(A+B_{m}^{u})C(t) \ \ \ \ \ \ \ (m=1, \dots, M)
\end{equation}
The corresponding connectivity graph of $B_{m}^{u}$ is denoted as
$G_m=(V_m, E_m)$. It is clear that for arbitrary $m$,
$V_m=V=\{|\phi_{1}\rangle,\dots,|\phi_{N}\rangle\}$. Hence, we can
write $G_m=(V, E_m)$ and define $G=(V, E)$ as the combined graph of
$G_m=(V, E_m)$ ($m=1, \dots, M$) where $E=E_1\cup E_2 \cup \dots
\cup E_M$. We then give the following theorem \cite{Dong VSC et al
2010}.
\begin{theorem}\label{generalcontrollable}
The system
\begin{equation}\label{generalmodel}
i\dot{C}(t)=(A+B^{u})C(t) \ \ \ \ \ \ C(0)=C_{0} \ \ \ \ \ \ B^{u}
\in \{B^{u}_{1}, \dots, B^{u}_{M}\}
\end{equation}
is pure state controllable if the controller $B^{u}$ can be
arbitrarily switched between controllers $B^{u}_{m}$ and $B^{u}_{n}$
($m \neq n$, $1\leq m, n \leq M$), and the
following assumptions hold: \\
(I) The combined graph $G$ for the system (\ref{generalmodel}) is connected. \\
(II) The combined graph $G$ remains connected after elimination of
edge pairs $(|\phi_{j}\rangle, $

$|\phi_{l}\rangle)$, $(|\phi_{a}\rangle, |\phi_{b}\rangle)$ such
that $\nu_{jl}= \nu_{ab}$ (degenerate transitions).
\end{theorem}

The proof of this theorem can be found in \cite{Dong VSC et al
2010}. The variable structure control strategy provides a useful
method for open-loop control design for a class of quantum systems
and is also applicable to qubit preparation for quantum information
processing (for details, see \cite{Dong VSC et al 2008}, \cite{Dong
VSC et al 2010}). Here, for open-loop control of quantum systems, we
only use the concept of variable structure control and consider
control inputs which are allowed to switch between a number of
different control structures. Moreover, note that sliding mode
control is an important aspect in variable structure control of
classical systems. Also, the sliding mode method can be used to deal
with uncertainties in the feedback control of quantum systems and
has potential applications to quantum state preparation and quantum
error correction \cite{Dong and Petersen 2009 NJP}.

\subsection{Incoherent control}
The above discussions on open-loop quantum control focus on coherent
control where we can manipulate the state of a quantum system by
applying semiclassical potentials in a fashion that preserves
quantum coherence \cite{Lloyd 2000}. The control inputs commonly
occur as tunable parameters in the Hamiltonian of the system and can
directly affect the coherent part of the dynamics. There exist some
physical situations where it is not possible or very difficult to
control the state of a quantum system directly with coherent
resources (unitary transformations) \cite{Romano and D'Alessandro
2006-1}. For example, it has been proven that a finite dimensional
open quantum system with Markovian dynamics is not controllable
using only coherent control \cite{Altafini 2003}, \cite{Altafini
2004}. A natural extension of these quantum control methods is to
introduce incoherent resources (called ``incoherent control") to
enhance our capability of controlling quantum systems or to help in
control design in order to accomplish specified tasks. The main
paradigms for introducing incoherent resources into open-loop
quantum control include incoherent control fields \cite{Shapiro and
Brumer 2003}, the use of auxiliary systems (environments) and
quantum measurements \cite{Vilela Mendes and Man'ko 2003}-\cite{Dong
et al 2008JCP}, \cite{Roa et al 2006}-\cite{ZhangM et al 2007}.

Quantum measurements are regarded as deleterious in accomplishing
coherent control tasks since they usually destroy the coherent
characteristics of the quantum system. However, recent results show
that quantum measurements can improve the controllability of quantum
systems in some situations. Indeed, sometimes a quantum system which
is not unitarily controllable can be controlled by the joint action
of projective measurement and unitary evolution \cite{Vilela Mendes
and Man'ko 2003}, \cite{Mandilara and Clark 2005}. Let us consider
the model (\ref{BLM1}), and let $G(\mathcal{A})$ denote the Lie
group generated by the operators $\{-iH_{0}, -iH_{1}, \dots,
-iH_{K}\}$. Also, let $O(N)$ be the $N$-dimensional orthogonal
group. We then have the following theorem \cite{Vilela Mendes and
Man'ko 2003}:

\begin{theorem}
Given any goal state $|\psi_{f}\rangle$, there is a family of
observables $M(|\psi_{f}\rangle)$ such that measurement of one of
these observables on any $|\psi_{0}\rangle$ plus unitary evolution
leads to $|\psi_{f}\rangle$ if $G(\mathcal{A})$ is equal to either
$O(N)$ or $sp(\frac{N}{2})$.
\end{theorem}

The proof of this theorem can be found in \cite{Vilela Mendes and
Man'ko 2003}. If $G(\mathcal{A})$ is equal to either $U(N)$ or
$sp(\frac{N}{2})$, the corresponding system is pure state
controllable and it might still be more efficient to use the
measurement plus evolution strategy \cite{Vilela Mendes and Man'ko
2003}. Several incoherent control schemes based on quantum
measurements have been proposed. For example, a control scheme for
mapping an unknown mixed quantum state onto a known pure state has
been proposed with the help of sequential measurements of two
noncommutative observables and without the use of unitary
transformations \cite{Roa et al 2006}. A probabilistic quantum
control strategy using indirect measurement is presented for the
remote control of quantum systems \cite{Mandilara and Clark 2005}.
Projective measurements on several identical initial states are used
to help complete the control design of quantum systems with pure
state controllable subspaces \cite{Dong et al 2007-1}. Projective
measurements are also combined with amplitude amplification for the
state control of locally controllable quantum systems \cite{Dong et
al 2008JCP}. In these incoherent control schemes, we emphasize the
role of measurement as a control tool. Although we may use the
measurement result, we present these methods as open-loop control
strategies since no real closed-loop is constructed in the control
process.

Quantum measurements can be used as an effective control tool in
quantum control. In practical applications, the realization of a
quantum measurement on a quantum system is usually accomplished by
entangling the quantum system with an auxiliary system (a probe).
Hence, one can also control a quantum system by controlling an
auxiliary system (e.g., a probe) entangled with the quantum system.
Under suitable conditions on the interaction of the system to be
controlled, the auxiliary system, and the environment, the system
dynamics is completely controllable by varying the initial state of
the auxiliary system \cite{Romano and D'Alessandro 2006-1},
\cite{Romano and D'Alessandro 2006-2} or manipulating the local
Hamiltonian of the auxiliary system \cite{Burgarth and Giovannetti
2007}, which has potential applications to quantum networks
\cite{Burgarth et al 2009}. Furthermore, if all kinds of coherent
and incoherent resources are available, in principle, we can control
a (closed or open) quantum system from an arbitrary initial state to
a predefined target state \cite{Kraus 1983}, \cite{Lloyd and Viola
2001}.

\section{Some results on closed-loop design methods}\label{Section4}
As presented in Section \ref{Section3}, open-loop control has
achieved many successes in the control of some simple quantum
systems. However, it suffers some difficulties for more complex
quantum control tasks such as suppressing decoherence and dealing
with disturbances in quantum systems. A natural solution to this
problem is to explore closed-loop control strategies. Two forms for
closed-loop control have been proposed for quantum systems:
closed-loop learning control and quantum feedback control. In
closed-loop learning control, each cycle of the closed-loop is
executed on a new sample. However, in quantum feedback control, the
same sample is involved during the whole process of control
\cite{Rabitz et al 2000}. In this section, we will present some
aspects of quantum feedback control after a brief introduction to
closed-loop learning control.

\subsection{Closed-loop learning control}
Closed-loop learning control has achieved great successes in the
control of laboratory chemical reactions \cite{Rabitz et al 2000},
\cite{Judson and Rabitz 1992}. The closed-loop learning control
procedure generally involves three elements \cite{Rabitz et al
2000}: (i) a trial laser control input design, (ii) the laboratory
generation of the control that is applied to the sample and
subsequently observed for its impact, and (iii) a learning algorithm
that considers the prior experiments and suggests the form of the
next control input. The initial trial control input may be a
well-designed laser pulse or a random control input. A feature of a
good closed-loop learning control strategy is its insensitivity to
the initial trials. An important task is to establish a good
learning algorithm for ensuring that the closed-loop learning
process converges to achieve a given control objective. Genetic
algorithms and several rapid convergence algorithms have been
employed for this task \cite{Judson and Rabitz 1992}, \cite{Zhu and
Rabitz 1998}. The control objective is usually formulated as an
optimal control problem by converting the problem into a problem of
optimizing a functional of the quantum states, control inputs,
control time, etc. In the learning process, an optimal control
problem is solved iteratively. First, one applies a trial input to a
sample to be controlled and observes the result. Second, a learning
algorithm suggests a better control input based on the prior
experiments. Third, one applies a ``better" control input to a new
sample. This process continues in order to achieve the control
objective. It is often easy to produce many identical-state samples
in laboratory chemical reactions. If the control objective is well
defined, there is a capability to apply specified control inputs to
the samples. Then, if a sufficiently intelligent learning algorithm
is applied to adjust the control inputs, this process will converge
to optimize the required objective and an optimal control will be
found \cite{Rabitz et al 2000}.

\subsection{Quantum feedback control}
As we know, feedback is an effective strategy in classical control
theory and the aim of feedback is to compensate for the effects of
unpredictable disturbances on a system under control, or to make
automatic control possible when the initial state of the system is
unknown. In classical control, many results have shown that feedback
control is superior to open-loop control. In feedback control, it is
usually necessary to obtain the information about the state of
system through measurement. However, measurements of a quantum
system will unavoidably disturb the state of measured quantum
system, which makes the situation more complex when applying
feedback to quantum systems. In spite of this difficulty, important
progress has been made and feedback has been used to improve the
control performance for squeezed states \cite{Thomsen et al 2002},
\cite{Haus and Yamamoto 1986}, quantum entanglement \cite{Yanagisawa
2006}, \cite{Yamamoto et al 2008}, \cite{Hill and Ralph 2008}, and
quantum state reduction \cite{van Handel et al 2005}-\cite{Combes
and Jacobs 2006}, \cite{Wiseman and Ralph 2006}-\cite{Griffith et al
2007} in many areas such as quantum optics \cite{Wiseman and Milburn
1993}, \cite{Wiseman and Milburn 2009}, superconducting quantum
systems \cite{Griffith et al 2007}, Bose-Einstein condensate
\cite{Wilson et al 2007} and nanomechanical systems \cite{Zhang et
al 2009 Nori}. Quantum feedback control has also been compared with
open-loop control (e.g., see \cite{Qi and Guo 2009}). It has been
proven for a specific quantum system that quantum feedback is
superior to open-loop control in dealing with uncertainties in
initial states and it has also been demonstrated via simulation that
feedback control is still better than open-loop control for dealing
with decoherence \cite{Qi and Guo 2009}.

In quantum feedback control, the two main approaches to information
acquisition are projective measurement and continuous weak
measurement. The system to be controlled is a quantum system,
however, the controller may be quantum, classical or a
quantum/classical hybrid. Several paradigms of quantum feedback have
been proposed, such as Markovian quantum feedback \cite{Wiseman and
Milburn 1993}, \cite{Wiseman 1994}, Bayesian quantum feedback
\cite{Doherty and Jacobs 1999}, \cite{Doherty et al 2000},
\cite{Doherty et al 2001}, and coherent quantum feedback \cite{Lloyd
2000}, \cite{James et al 2008}. In Markovian quantum feedback, any
time delay is ignored and a memoryless controller is assumed. That
is, the measurement record is immediately fed back onto the system
to alter the system dynamics and may then be forgotten \cite{Wiseman
et al 2002}. Hence, the equation describing the resulting evolution
is a Markovian master equation. In Bayesian quantum feedback, the
process is divided into two steps involving state estimation and
feedback control. The best estimates of the dynamical variables are
obtained continuously from the measurement record, and fed back to
control the system dynamics \cite{Doherty and Jacobs 1999}. The
dynamical equation describing the resulting evolution is
non-Markovian. In coherent quantum feedback, the feedback controller
itself is a quantum system, and it processes quantum information.
This is greatly different from Markovian and Bayesian quantum
feedback where the feedback information from measurement results is
classical information and the feedback controller is a classical
controller. In the following, we will present several important
aspects of quantum feedback control including quantum filtering,
feedback stabilization, optimal feedback control and robust control.

\subsubsection{Identification and filtering/estimation}
To design an effective quantum feedback control system, it is
necessary to obtain knowledge of a system model and system states.
This can be regarded as a quantum system identification or state
estimation problem. The simplest example is to estimate the system
state for feedback through projective measurements on a quantum
ensemble. This is a simple problem in classical control which can
usually be accomplished through direct measurements. However it
becomes much more complex due to the quantum collapse postulate of
measurement in quantum control. This problem of identifying quantum
states is also called quantum state tomography in quantum
information and some procedures have been developed to
experimentally determine an unknown quantum state \cite{Nielsen and
Chuang 2000}. Some of these procedures can be used in quantum
feedback control. Another example is the identification of
parameters for a quantum operation (e.g., a quantum gate or the
system Hamiltonian). This procedure is called as quantum process
tomography and some techniques have been developed (e.g., see
\cite{Geremia and Rabitz 2002}, \cite{Nielsen and Chuang 2000}). To
acquire information about a quantum operation, an attractive
approach is quantum nondemolition measurements which do not disturb
the value to be measured. However, it is not practical to design
quantum nondemolition measurements for most problems (see
\cite{Braginsky and Khalili 1996} for details).

Furthermore, it is often useful to estimate an unknown dynamical
parameter through continuously monitoring an open quantum system
\cite{Mabuchi 1996}, \cite{Gambetta and Wiseman 2001}. Since
continuous observations of open quantum systems are inherently
noisy, quantum filtering theory is required for extracting
information from a noisy signal \cite{Bouten et al 2007}. Quantum
filtering theory has been developed for quantum optical systems and
is used as an essential component in some quantum feedback control
strategies \cite{Belavkin 1999}, \cite{Bouten et al 2008},
\cite{Belavkin 1992}, \cite{James 2004}. In the framework of
noncommutative (or quantum) probability theory, a broad class of
quantum stochastic differential equations can be obtained (for
details, see, e.g., \cite{Bouten et al 2007}). In fact, the SME
model (\ref{SME1}) is a quantum filtering equation which describes
the evolution of a conditional state. Recently, several robustness
properties of quantum filtering and estimation have also been
investigated for linear and nonlinear quantum systems \cite{Stockton
et al 2004}, \cite{Yamamoto 2006}, \cite{Yamamoto and Bouten 2009}.

\subsubsection{Feedback stabilization/control}
Based on the use of feedback information from measurements or
estimation, one can design a feedback controller to control or
stabilize a quantum system. In Markovian quantum feedback, the
feedback Hamiltonian is commonly selected as a linear function of
the feedback signal. For example, the feedback Hamiltonian in
Markovian feedback control via homodyne detection may be selected as
$H_{fb}(t)=F(t)I(t)$, where $F(t)$ is a Hermitian operator and
$I(t)$ is a current from the measurement \cite{Wiseman and Milburn
1993}, \cite{Wiseman et al 2002}, \cite{Wang and Wiseman 2001}. Then
the feedback Hamiltonian will enter the system Hamiltonian to alter
the system evolution. Markovian quantum feedback has been widely
applied in the stabilization of arbitrary pure states of two-level
systems \cite{Wang and Wiseman 2001}, noiseless subspace generation
\cite{Ticozzi and Viola 2008}, continuous quantum error correction
\cite{Ahn et al 2003}, etc. In Bayesian quantum feedback, the
feedback Hamiltonian is a general function of the measurement record
\cite{Doherty and Jacobs 1999}, \cite{Ruskov and Korotkov 2002},
which is used to control the system dynamics. Bayesian feedback is
usually superior to Markovian feedback since it uses more
information. However, it is more difficult to implement Bayesian
feedback than Markovian feedback due to the existence of the
estimation step \cite{Wiseman et al 2002}. Bayesian feedback has
also been applied in the preparation of quantum states \cite{Ruskov
and Korotkov 2002} and quantum error correction \cite{Ahn et al
2002}, \cite{Sarovar et al 2004}. If the feedback delay cannot be
ignored, the feedback Hamiltonian must include a delay parameter
\cite{Giovannetti et al 1999} which in some cases has a
qualitatively similar effect to that of unprecise measurement
\cite{Wang et al 2001b}. Both Markovian feedback and Bayesian
feedback use information from measurement (or estimation) and the
feedback controller is usually a classical controller. Another
different feedback paradigm is coherent quantum feedback, where a
quantum controller (i.e., the controller is another quantum system)
directly affects the system dynamics to be controlled \cite{Lloyd
2000}, \cite{Mabuchi 2008}, \cite{Nurdin et al 2009}.

In quantum feedback control, the filtering SME is an important
dynamical equation for the design of the feedback control system.
For example, for a quantum system described by an SME, one may use
stochastic Lyapunov techniques to design a feedback control law for
asymptotically stabilizing the quantum system to an objective state.
For the model (\ref{SME2}), if $\mathcal{S}=\{\rho \in
\mathbb{C}^{N\times N}: \rho=\rho^{\dag}, \text{tr}{\rho}=1,
\rho\geq 0\}$, we have the following result \cite{Mirrahimi and van
Handel 2007}:

\begin{theorem}
Consider the system (\ref{SME2}) evolving in the set $\mathcal{S}$.
Let the final state $\rho_{f}=v_{f}v^{\dag}_{f}$, where $v_{f}$ is
an eigenstate $|\phi_{m}\rangle$ ($1\leq m \leq N$), and let $\gamma
>0$. Consider the following control law:
\begin{enumerate}
    \item
    $u_{t}=-\text{tr}(i[F_{y}, \rho_{t}]\rho_{f})$ if $\text{tr}(\rho_{t}\rho_{f})\geq
    \gamma$.
    \item $u_{t}=1$ if $\text{tr}(\rho_{t}\rho_{f})\leq
    \gamma/2$.
    \item If $\rho_{t} \in \mathcal{B}=\{\rho: \gamma/2 < \text{tr}(\rho\rho_{f})< \gamma\}$,
    then $u_{t}=-\text{tr}(i[F_{y}, \rho_{t}]\rho_{f})$ if $\rho_{t}$ last entered $\mathcal{B}$ through the boundary
    $\text{tr}(\rho\rho_{f})=\gamma$, and $u_{t}=1$ otherwise.
\end{enumerate}
Then there exists a $\gamma >0$ such that $u_{t}$ globally
stabilizes (\ref{SME2}) around $\rho_{f}$ and
$\mathbb{E}\rho_{t}\rightarrow \rho_{f}$ as $t \rightarrow \infty$.
\end{theorem}

\subsubsection{LQG control}
Under appropriate assumptions, some quantum optical systems can be
approximately modeled by LQSDEs driven by quantum Wiener processes
\cite{Gardiner and Zoller 2000}, \cite{Wiseman and Doherty LQG}.
This simplification provides an opportunity to develop quantum
linear-quadratic-Gaussian (LQG) control to obtain an optimal
feedback strategy \cite{Belavkin 1983}, \cite{Mancini and Wiseman
2007}, \cite{Belavkin 1999}, \cite{Doherty and Jacobs 1999},
\cite{Doherty et al 2000}, \cite{Wiseman and Doherty LQG},
\cite{Zhang 2008}, \cite{Shaiju et al 2008}. In LQG control, the
goal is to find an optimal feedback control law for a stochastic
linear system by optimizing a quadratic cost functional. In some
cases, results from classical LQG control can be applied to quantum
systems \cite{Sayed Hassen et al 2009}. An important result is the
separation theorem which applies when the system is linear, the cost
functional is quadratic in the system variables and the noises are
Gaussian. In this situation, the values of optimal estimates are fed
back. Then, when calculating the feedback required for the optimal
control, we may assume that the dynamical variables are known
exactly \cite{Doherty and Jacobs 1999}.

In the quantum LQG control problem, the optimal control is also a
linear feedback control law. The controller may be a classical
controller \cite{Shaiju et al 2008} or a quantum controller
\cite{Mabuchi 2008}, \cite{Nurdin et al 2009}. For classical
controllers, it is straightforward to apply results in classical LQG
control to quantum LQG problems. In fact, some results on quantum
LQG control with classical controllers have been developed in
\cite{Mancini and Wiseman 2007}, \cite{Doherty and Jacobs 1999},
\cite{Doherty et al 2000}, \cite{Wiseman and Doherty LQG},
\cite{Shaiju et al 2008}. For controllers which are themselves
quantum systems, we must add some additional constraints on
coefficient matrices to ensure that the controller is physically
realizable. For example, \cite{James et al 2008} has developed a
notion of physical realizability based on the concept of an open
quantum harmonic oscillator as the basic components of a physically
realizable quantum system. Under several reasonable assumptions on a
system of the form (\ref{LQSE}) (e.g., $n_{y}$ is even,
$n_{\omega}\geq n_{y}$, $F_{\omega}=I+i\text{diag}(J, \dots, J)$,
etc.; see \cite{James et al 2008} for details), we have the
following results on physical realizability \cite{James et al 2008}:

\begin{theorem}\label{physicalRealization}
The system (\ref{LQSE}) is physically realizable if and only if:
\begin{equation}
iA\Theta +i\Theta A^{T}+BT_{\omega}B^{T}=0,
\end{equation}
\begin{equation}
B\begin{bmatrix} I_{n_{y}\times n_{y}} \\
0_{(n_{\omega}-n_{y})\times n_{y}}
\end{bmatrix}=\Theta C^{T} \text{diag}_{N_{y}}(J)
\end{equation}
\begin{equation}
D=\begin{bmatrix} I_{n_{y}\times n_{y}} && 0_{n_{y} \times
(n_{\omega}-n_{y})}
\end{bmatrix}
\end{equation}
where $N_{y}=\frac{n_{y}}{2}$ and
$T_{\omega}=\frac{1}{2}(F_{\omega}-F_{\omega}^{T})$.
\end{theorem}

Hence, the problem of quantum LQG control with a quantum controller
is converted into the following problem: Given a canonical
commutation relation, find a feedback controller, whose coefficient
matrices satisfy the conditions of physical realizability in Theorem
\ref{physicalRealization}, that minimizes a quadratic performance
index. The additional constraints on the coefficients for
controllers make the problem more complex. Indeed, new algorithms
are required to be developed in order to solve this problem. In
\cite{Nurdin et al 2009}, a numerical procedure based on an
alternating projection algorithm has been developed to solve this
control problem. Furthermore, several results on synthesizing
coherent quantum controllers have also been presented (e.g., see
\cite{Nurdin et al 2009 SIAM}).

\subsubsection{Robust control}
In practical applications, it is unavoidable that quantum systems
are subject to all kinds of disturbances and uncertainties
\cite{Zhang and Rabitz 1994}. Many instances of incomplete knowledge
and unknown errors can also be treated as uncertainties. Hence,
robustness has been identified as an important aspect for the
practical application of quantum technology \cite{D'Helon and James
2006}, \cite{Bacon et al 1999}, \cite{Rabitz 2002}. Recently,
several robust control methods in classical control theory have been
extended into quantum domain. For example, the small gain theorem
has been applied to the stability analysis of quantum feedback
networks \cite{D'Helon and James 2006}. A transfer function approach
has been applied to the feedback and robust control for single input
single output quantum systems \cite{Yanagisawa and Kimura 2003-1},
\cite{Yanagisawa and Kimura 2003-2}. An $H^{\infty}$ controller
synthesis problem has been formulated for a class of linear quantum
stochastic systems \cite{James et al 2008}. A sliding mode control
approach has been applied to robust control design for quantum
systems where bounded uncertainties exist in the system Hamiltonian
\cite{Dong and Petersen 2009 NJP}. Some open-loop and feedback
approaches to quantum control can lead to a certain degree of
robustness. Since feedback plays a key role in robust control, in
this survey we present robust control within the framework of
quantum feedback control. In the following, we will only present
several specific results on $H^{\infty}$ control for quantum
systems, which is based on the reference \cite{James et al 2008}.

Consider the quantum system to be controlled to be described by the
following LQSDE defined in an analogous way to (\ref{LQSE}):
\begin{equation}\label{LQSEplant}
\begin{array}{l}
dx(t)=Ax(t)dt+[B_{0}\ \ B_{1}\ \ B_{2}]\times [dv(t)^{T}\ \ dw(t)^{T}\ \ du(t)^{T}]^{T}; \ x(0)=x_{0} \\
dz(t)=C_{1}x(t)dt+D_{12}du(t)\\
dy(t)=C_{2}x(t)dt+[D_{20}\ \ D_{21}\ \ 0_{n_{y}\times n_{u}}]\times
[dv(t)^{T}\ \ dw(t)^{T}\ \ du(t)^{T}]^{T}.
\end{array}
\end{equation}
where the input $w(t)$ represents a disturbance signal of the form
(\ref{noiseinput}). Also, $dv(t)$ represents any additional quantum
noise in the system. The control input $u(t)$ can be written as
\begin{equation}
du(t)=\beta_{u}(t)dt+d\tilde{u}(t)
\end{equation}
where $\beta_{u}(t)$ is the adapted, self-adjoint part of $u(t)$ and
$\tilde{u}(t)$ is the noise part of $u(t)$. The vectors $v(t)$,
$\tilde{w}(t)$, and $\tilde{u}(t)$ are quantum noises with positive
semidefinite Hermitian It\^{o} matrices $F_{v}$, $F_{\tilde{w}}$ and
$F_{\tilde{u}}$.

The controller is assumed to be of the following form
\begin{equation}\label{LQSEcontroller}
\begin{array}{l}
dx(t)=A_{K}\xi(t)dt+[B_{K1}\ \ B_{K}]\times [dv_{K}(t)^{T}\ \ dy(t)^{T}]^{T}; \ \xi(0)=\xi_{0} \\
du(t)=C_{K}\xi(t)dt+[B_{K0}\ \ 0_{n_{u}\times n_{y}}]\times
[dv_{K}(t)^{T}\ \ dy(t)^{T}]^{T}
\end{array}
\end{equation}
where $\xi(t)=[\xi_{1}(t) \dots \xi_{n}(t)]^{T}$ is a vector of
self-adjoint controller variables. The noise $v_{K}(t)=[v_{K1}(t)\
\dots v_{KK_{v}}(t)]^{T}$ is a vector of noncommutative Wiener
processes with canonical Hermitian It\^{o} matrix $F_{v_{K}}$.

By identifying $\beta_{u}=C_{K}\xi(t)$ and interconnecting
(\ref{LQSEplant}) and (\ref{LQSEcontroller}), we can obtain the
resulting closed-loop system (see \cite{James et al 2008}). The goal
of the $H^{\infty}$ controller synthesis problem is to find a
controller (\ref{LQSEcontroller}) for a given disturbance
attenuation parameter $g>0$ such that the resulting closed-loop
system satisfies (for details, see \cite{James et al 2008})
\begin{equation}
\int_{0}^{t}\langle\beta_{z}(s)^{T}\beta_{z}(s)+\epsilon
\eta(s)^{T}\eta(s)\rangle ds\leq
(g^{2}-\epsilon)\int_{0}^{t}\langle\beta_{\omega}(s)^{T}\beta_{\omega}(s)\rangle
ds+\mu_{1}+\mu_{2}t, \forall t>0
\end{equation}
for some real constants $\epsilon$, $\mu_{1}$, $\mu_{2}>0$, where
$\beta_{z}(t)=[C_{1}\ \ D_{12}C_{K}][x(t)^{T}\ \ \xi(t)^{T}]^{T}$.

Furthermore, we suppose that the system (\ref{LQSEplant}) satisfies
the following assumption (Assumption A): 1)
$D_{12}^{T}D_{12}=E_{1}>0$; 2) $D_{21}D_{21}^{T}=E_{2}>0$; 3) The
matrix $[\begin{smallmatrix}A-i\omega I && B_{2}
\\ C_{1} && D_{12}\end{smallmatrix}]$ is full rank for all $\omega\geq
0$; 4) The matrix $[\begin{smallmatrix}A-i\omega I && B_{1}
\\ C_{2} && D_{21}\end{smallmatrix}]$ is full rank for all $\omega\geq
0$.

The results on quantum $H^{\infty}$ control will be stated in terms
of the following pair of algebraic Riccati equations:
\begin{equation}\label{Riccati1}
\begin{array}{l}
(A-B_{2}E_{1}^{-1}D_{12}^{T}C_{1})^{T}X+X(A-B_{2}E_{1}^{-1}D_{12}^{T}C_{1})
+X(B_{1}B_{1}^{T}-g^{2}B_{2}E_{1}^{-1}B_{2}^{T})X\\
+g^{-2}C_{1}^{T}(I-D_{12}E_{1}^{-1}D_{12}^{T})C_{1}=0
\end{array},
\end{equation}
\begin{equation}\label{Riccati2}
\begin{array}{l}
(A-B_{1}D_{21}^{T}E_{2}^{-1}C_{2})Y+Y(A-B_{1}D_{21}^{T}E_{2}^{-1}C_{2})^{T}
+Y(g^{-2}C_{1}^{T}C_{1}-C_{2}^{T}E_{2}^{-1}C_{2})Y\\
+B_{1}(I-D_{21}^{T}E_{2}^{-1}D_{21})B_{1}^{T}=0
\end{array}
\end{equation}
where $X$ and $Y$ are positive definite symmetric matrices. The
solutions to these Riccati equations will be required to satisfy the
following assumption (Assumption B): 1)
$A-B_{2}E_{1}^{-1}D_{12}^{T}C_{1}
+(B_{1}B_{1}^{T}-g^{2}B_{2}E_{1}^{-1}B_{2}^{T})X$ is a stability
matrix; 2) $A-B_{1}D_{21}^{T}E_{2}^{-1}C_{2}
+Y(g^{-2}C_{1}^{T}C_{1}-C_{2}^{T}E_{2}^{-1}C_{2})$ is a stability
matrix; 3) The matrix $XY$ has a spectral radius strictly less than
one.

If the Riccati equations (\ref{Riccati1}) and (\ref{Riccati2}) have
solutions satisfying Assumption B, a controller of the form
(\ref{LQSEcontroller}) will solve the $H^{\infty}$ control problem
under consideration and the system matrices of the controller can be
constructed from the Riccati solutions as follows:
\begin{equation}\label{Controller}
\begin{array}{l}
A_{K}=A+B_{2}C_{K}-B_{K}C_{2}+(B_{1}-B_{K}D_{21})B_{1}^{T}X \\
B_{K}=(I-YX)^{-1}(YC_{2}^{T}+B_{1}D_{21}^{T})E_{2}^{-1}\\
C_{K}=-E_{1}^{-1}(g^{2}B_{2}^{T}X+D_{12}^{T}C_{1}).
\end{array}
\end{equation}

The necessary and sufficient conditions on $H^{\infty}$ controller
synthesis can be described as follows:
\begin{theorem}
Necessity. Consider the system (\ref{LQSEplant}) and suppose that
Assumption A is satisfied. If there exists a controller of the form
(\ref{LQSEcontroller}) such that the resulting closed-loop system is
strictly bounded real with disturbance attenuation $g$, then the
Riccati equations (\ref{Riccati1}) and (\ref{Riccati2}) will have
stabilizing solutions $X\geq 0$ and $Y\geq 0$ satisfying Assumption
B.

Sufficiency. Suppose the Riccati equations (\ref{Riccati1}) and
(\ref{Riccati2}) have stabilizing solutions $X\geq 0$ and $Y\geq 0$
satisfying Assumption B. If the controller (\ref{LQSEcontroller}) is
such that the matrices $A_{K}$, $B_{K}$, $C_{K}$ are as defined in
(\ref{Controller}), then the resulting closed-loop system will be
strictly bounded real with disturbance attenuation $g$.
\end{theorem}

It is worth noting that when the controller which is constructed via
this approach is a quantum controller, we need to take into account
the physical realizability conditions using Theorem
\ref{physicalRealization}.

\section{Conclusions and perspectives}\label{Section5}
In this paper, we have surveyed developments in quantum control
theory and its applications from a control systems perspective. Some
results on controllability of quantum systems, open-loop control
strategies and closed-loop design methods have been presented. Also,
some applications of quantum control theory are briefly mentioned in
this survey. Although great progress has been made in this area,
quantum control is still in its infancy. Moreover, there exist
several essential differences between quantum control theory and
classical control theory. For example, controlling quantum
entanglement and protecting quantum coherence are important two
classes of quantum control tasks. However, no corresponding tasks
exist in classical control theory. In classical feedback control,
measurement is taken as the main tool of acquiring feedback
information to deal with uncertainties and it is assumed that
measurement does not affect the measured system. However,
measurement in quantum feedback control unavoidably introduces
essential uncertainties when we acquire feedback information through
measurement to deal with uncertainties of quantum control systems.
Hence, it is necessary to develop more new theory and approaches to
control quantum phenomena. We now outline several important topics
that are worth exploring, emphasizing different aspects of quantum
control (a more detailed discussion on mathematical and algorithmic
challenges in quantum control can be found in \cite{Brown and Rabitz
2002}).

(1) Quantum incoherent control. Quantum coherent control is a
powerful control paradigm in open-loop quantum control and many
important results have been achieved. However, it may not be
practical to accomplish some quantum control tasks using only
coherent control approaches. Recently, incoherent resources have
been proven useful for accomplishing some quantum control tasks. An
example of this is the fact that quantum measurement can serve as a
useful control tool. However, there are only few results on the use
of quantum incoherent control. Hence, it is desirable to develop
systematic and new approaches to construct quantum incoherent (or
coherent/incoherent hybrid) controllers which are easily physically
realized \cite{Pechen and Rabitz 2006}.

(2) Quantum feedback control. Feedback plays a key role in classical
control theory. It is desirable to develop systematic feedback
control methods for emerging quantum engineering applications.
Although many results on quantum feedback control have been
presented, they are usually constrained to some special cases (e.g.,
linear feedback, Markovian approximation, simple quantum systems).
Many fundamental problems still need to be considered: How does the
weak measurement \cite{Lloyd and Slotine 2000} affect quantum
systems in quantum feedback control? How does one design feedback
controllers for nonlinear quantum systems \cite{Jacobs and Lund
2007} and is it possible to control nonlinear dynamics of quantum
systems using linear controllers? What approaches are effective for
control of quantum systems with non-Markovian dynamics
\cite{Giovannetti et al 1999}, \cite{Wang et al 2001b}, \cite{Cui et
al 2008}? What is the capability of feedback to deal with
uncertainties in quantum systems? How do we synthesize complex
quantum feedback network systems \cite{Gough and James 2009}?

(3) Robust control of quantum systems. It is inevitable that quantum
systems are subject to many kinds of uncertainties (including
disturbances and noises). Hence, the requirement of robustness in
the presence of uncertainties has been recognized as one of the key
tasks for developing practical quantum technologies. Existing
results in this area focus mainly on quantum systems whose dynamics
can be described by linear equations in the Heisenberg picture
\cite{James et al 2008}, \cite{Maalouf and Petersen 2009} or whose
uncertainties can be approximately described as a perturbation in
the system Hamiltonian. It is desirable to develop systematic robust
control approaches to deal with more general kinds of uncertainties
existing in practical quantum systems.

(4) Decoherence control. Decoherence is the main obstacle to the
application of quantum information technology. Many quantum control
systems are essentially open quantum systems \cite{Breuer and
Petruccione 2002}, \cite{Carmichael 1993} and interaction with
external environments (including external controls and the
measurement apparatus) unavoidably introduces decoherence
\cite{ZhangJ et al 2007}. A typical example of decoherence control
is quantum error correction \cite{Ahn et al 2002}, \cite{Sarovar et
al 2004}, \cite{Ahn et al 2003}. Several decoherence control
approaches have been proposed. For example, ``bang-bang" control has
been proposed for dynamic decoherence suppression in two-state
quantum systems \cite{Viola and Lloyd 1998}. A classical feedback
strategy has been proposed to eliminate decoherence in open quantum
systems \cite{Ganesan and Tarn 2007}. Optimal control techniques
have been applied to decoherence control of Markovian and
non-Markovian open quantum systems \cite{Cui et al 2008},
\cite{ZhangJ et al 2005}, \cite{Gordon and Kurizki 2008}. In spite
of these efforts, there are still many challenges for decoherence
control and the development of systematic and practical decoherence
control methods is still an important problem.

(5) Entanglement control. Quantum entanglement is the most important
resource in quantum information technology. Several optimal control
and feedback control approaches have been proposed to control
quantum entanglement \cite{Yamamoto et al 2007}, \cite{Mancini and
Wiseman 2007}, \cite{Yanagisawa 2006}, \cite{Yamamoto et al 2008},
\cite{Hill and Ralph 2008}, \cite{Zhang et al 2010}, \cite{Cui et al
2009}. However, existing results are only a first step in
entanglement control. Quantum entanglement has many unique
characteristics which have no classical counterparts. Hence, it is
desirable to develop completely new methods to control quantum
entanglement. Moreover, we still lack a suitable theory that can
effectively characterize quantum entanglement. This makes the
development of a control theory for quantum entanglement more
challenging.



\section*{Acknowledgment}
The authors thank the editor James Lam for inviting them to write
this survey. The authors would like to thank the referees and K
Jacobs, M R James, A I Maalouf, H I Nurdin, A Pechen, B Qi, H
Rabitz, S Z Sayed Hassen, T J Tarn, I Vladimirov, R B Wu, M
Yanagisawa and G F Zhang for their helpful comments and suggestions.
The authors also thank D Burgarth, V Giovannetti and F Nori for
pointing out several useful references.

%

\end{document}